\documentclass[aps,prb,twocolumn,longbibliography]{revtex4-1}
\usepackage{makeidx}
\usepackage{amsmath,amssymb,amsfonts,amsthm}
\usepackage{graphicx,bm}
\usepackage{hyperref}

\begin{document}

\title{Probing ferroelectricity in highly conducting materials through their
elastic response: persistence of ferroelectricity in metallic BaTiO$%
_{3-\delta }$}
\date{}
\author{F. Cordero,$^{1}$ F. Trequattrini,$^{2,1}$ F. Craciun,$^{1}$ H. T.
Langhammer$^{3}$, D.A.B. Quiroga$^{4}$ and P.S. Silva Jr$^{4}$}
\affiliation{$^{1}$ Istituto di Struttura della Materia-CNR (ISM-CNR), Area della Ricerca
di Roma - Tor Vergata,\\
Via del Fosso del Cavaliere 100, I-00133 Roma, Italy}
\affiliation{$^{2}$ Dipartimento di Fisica, Universit\`{a} di Roma \textquotedblleft La
Sapienza\textquotedblright , P.le A. Moro 2, I-00185 Roma, Italy}
\affiliation{$^{3}$ Institute of Chemistry, Martin Luther University Halle-Wittenberg,
Kurt-Mothes-Str. 2, 06120 Halle (Saale), Germany}
\affiliation{$^{4}$ Department of Physics, Federal University of S\~{a}o Carlos, P.O. Box
676, 13565-905 S\~{a}o Carlos, SP, Brazil}

\begin{abstract}
The question whether ferroelectricity (FE) may coexist with a metallic or
highly conducting state, or rather it must be suppressed by the screening
from the free charges, is the focus of a rapidly increasing number of
theoretical studies and is finally receiving positive experimental
responses. The issue is closely related to the thermoelectric and
multiferroic (also magnetic) applications of FE materials, where the
electrical conductivity is required or spurious. In these circumstances, the
traditional methods for probing ferroelectricity are hampered or made
totally ineffective by the free charges, which screen the polar response to
an external electric field. This fact may explain why more than 40 years
passed between the first proposals of FE metals and the present experimental
and theoretical activity. The measurement of the elastic moduli, Young's
modulus in the present case, versus temperature is an effective method for
studying the influence of doping on a FE transition because the elastic
properties are unaffected by electrical conductivity. In this manner, it is
shown that the FE transitions of BaTiO$_{3-\delta }$ are not suppressed by
electron doping through O vacancies; only the onset temperatures are
depressed, but the magnitudes of the softenings, and hence of the
piezoelectric activity, are initially even increased.
\end{abstract}

\maketitle


\section{Introduction}

\bigskip The concept of ferroelectric metal is rather counterintuitive,
since a material is defined ferroelectric (FE) if it has a spontaneous
polarization that can be switched by an external electric field, but within
a metal an external electric field is screened out by the free charges, so
that in this context the term 'ferroelectric' is sometimes put in quotes, or
substituted by 'polar'. Nonetheless, non-centrosymmetric displacements of
atoms with different charges are possible also in a metal and would produce
a local polarization. The first mentions to ferroelectric metals regarded V$%
_{3}$Si\cite{AB65} and Na$_{x}$WO$_{3}$.\cite{RSJ64} The first is one of the
A-15 compounds, which become superconductors close to a structural
instability, whose order parameter was initially proposed to be a soft
transverse optical phonon, as for FE transitions.\cite{AB65} It has later
been shown that the structural transition was driven by electronic band
splitting,\cite{Pyt70b} rather than a FE-like soft phonon. Also in
tungsten-bronze (TTB) Na$_{x}$WO$_{3}$, which becomes superconducting in a
doping range around $x\sim 0.3$, it was suggested that ferroelectric-like
distortions are present and might favor superconductivity. Though other TTBs
are ferroelectric,\cite{ZFS15} FE displacements have not been confirmed by
diffraction experiments in Na$_{x}$WO$_{3}$, but in WO$_{3}$, corresponding
to $x=0$, the W atoms are antiferroelectrically displaced.\cite{LSS99d} On
the other hand, x-ray and electron diffraction studies on TTB fluorides
suggest that ferroelectricity is present in a broader range of temperature
and compositions both in TTB fluorides and oxides.\cite{FMR04} After these
early suggestions,\cite{RSJ64,AB65,BBC71} the possibility of coexistence of
FE and metallic states received no further consideration, except marginally
in the context of high $T_{\mathrm{c}}$ superconductivity,\cite{CCC92,BW01}
until the phenomenon has been proposed to occur in the pyrochlore Cd$_{2}$Re$%
_{2}$O$_{7}$,\cite{SKM04} which is superconducting below 1~K, after
analyzing the elastic softening with considerations similar to those of
Anderson and Blount.\cite{AB65} Again, no evidence of spontaneous
polarization has been found, and the observations may be better explained in
terms of a low-temperature piezoelectric but non-ferroelectric phase.\cite%
{SKM04,II10}

More solid evidence of coexistence of ferroelectricity and metallic
conduction has been provided by heavily doping BaTiO$_{3-\delta }$ with O
vacancies (V$_{\text{O}}$), and showing with a combination of several
experimental techniques that the originally FE transition remains with
essentially the same characteristics after doping also in the metallic state,%
\cite{KTK10} even though with reduced $T_{\mathrm{C}}$. It was suggested
that the coexistence of metallic and FE states is possible, until the
Thomas-Fermi screening length does not exceed the FE correlation length.\cite%
{KTK10} Again, the interpretation of the persistence of the FE transition in
the metallic state of BaTiO$_{3-\delta }$ has been criticized, on the basis
that there would be a phase separation into highly doped metallic and
undoped FE phases,\cite{JLJ11} so that the demonstration by neutron
diffraction that the low temperature phase of LiOsO$_{3}$ is isostructural
with LiNbO$_{3}$, a well known FE, could be considered the first clear
evidence of coexisting metallic and FE states.\cite{SGW13} Very recently,
also the metallic layered perovskite Ca$_{3}$Ru$_{2}$O$_{7}$ has been shown
by optical second harmonic generation (SHG) and TEM to have polar domains
which, being also ferroelastic, can be switched by mechanical stress.\cite%
{LGP18} The origin of the polar displacements in Ca$_{3}$Ru$_{2}$O$_{7}$,
rather than to the usual mechanisms in ferroelectrics, would be due to a
trilinear coupling involving the polar displacements and two types of
tilting of the RuO$_{6}$ octahedra,\cite{LGP18} a mechanism called hybrid
improper ferroelectricity.\cite{BF11} Another class of polar metals includes
few transition metal monopnictides such as TaAs and NbAs,\cite{WPM17} where
the SHG coefficient is extremely high along the polar axis. It is proposed
that this is due to the particular topology of the Fermi surface,
characteristic of a Weyl-semimetal,\cite{WPM17} but also a simple model of
uniaxial polar semiconductor can explain the observations.\cite{PWL18}

To our knowledge, further examples of FE metals have been announced in thin
films, due to the substrate influence\cite{KPY16} or the extremely small
thickness that allows the external field to penetrate and switch the
polarization,\cite{FZP18} but not in other bulk materials, and the brief
review of the previous attempts at proving their existence demonstrates that
this is not an easy task. Yet, FE metals would not just be rare curiosities,
but might find technological applications based, for example, on giant
nonlinear optical phenomena.\cite{WPM17} New theoretical and computational
studies on polar metals are appearing, not only dealing with new mechanisms
that induce polar displacements like improper hybrid ferroelectricity\cite%
{BF11} or Weyl-semimetals,\cite{WPM17} but also considering the effect of
electric conductivity on the traditional lattice and electrostatic
mechanisms that induce the FE instability.\cite{GC14,BB16,LGA16} The
prevailing opinion, that the polar distortions cannot develop in a metal due
to electrostatic screening,\cite{LG77,Coh92} is evolving. It has been
estimated that the polar distortions cannot survive electron doping in BaTiO$%
_{3}$, but may do it in PbTiO$_{3}$\cite{HJ16} and the hypothetical (at that
time\cite{ASC18}) SnTiO$_{3}$,\cite{MHJ17} thanks to the local influence of
the lone pairs of Pb$^{2+}$ and Sn$^{2+}$. Afterwards, it has been
calculated that the FE instability in BaTiO$_{3}$ is so short-ranged, that
it can persist at high electron densities,\cite{WLB12} as previously
suggested,\cite{KTK10} and the critical electron density for suppressing FE
has been calculated as 0.085 electrons/cell,\cite{ISM12} much higher than in
the previous\cite{KTK10} and present investigations revealing the
persistence of FE upon doping. It has also been estimated that the charged
impurities are more detrimental than free electrons to FE.\cite{CGN13} More
recently, it has been proposed that the electron screening may even favor
the polar distortions under certain circumstances.\cite{ZFE18}

In addition to new FE metals, with their promising but yet unexplored
implications and applications, highly doped classic ferroelectrics already
find practical applications as thermoelectric materials.\cite{LYW09} Other
examples of materials where ferroelectricity is generally accompanied by
high electrical conductivity are the multiferroics,\cite{DLC15} often
obtained by doping magnetic ions in FE materials.\cite{CCF16} In all these
cases, the experimental verification of the coexistence of ferroelectricity
with a highly conducting or metallic state is a major problem, since the
traditional tools for probing ferroelectricity, which are directly sensitive
to the polarization and require or cause sample poling, are made ineffective
by the free charge screening. In these cases one must resort to indirect
methods, like the determination of the cell structure by diffraction or the
nonlinear optic response by Second Harmonic Generation (SHG).

Recently, it has been shown that the temperature dependence of the elastic
moduli across a FE transition may be used to probe the piezoelectric
response even in unpoled ceramics,\cite{CCT16,Cor18,Cor18c} and, since the
elastic response is insensitive to free charges, this seems the ideal tool
for the study of coexisting FE and conducting or metallic states. As an
example, it will be shown that in this manner it is possible, with
relatively little experimental effort, to confirm the persistence of the FE
transition in BaTiO$_{3-\delta }$ doped with V$_{\text{O}}$, as initially
found by Kolodiazhnyi \textit{et al.,}\cite{KTK10} and that, in spite of the
depression of $T_{\mathrm{C}}$, the piezoelectric coupling may be even
enhanced, possibly as suggested in Ref. \onlinecite{ZFE18}.

\section{Experimental}

The measurements were made on BaTiO$_{3}$ samples prepared in two different
laboratories, which we label \#1 and \#2. Samples \#1.1 and \#1.2, with
dimensions $42\times 6.3\times 0.68$~mm$^{3}$, were cut from a same bar of
BaTiO$_{3}$ (BT) prepared in the Department of Physics of UFSCar-Brazil,
starting from commercial high purity powder of barium titanate (IV) (99.9\%,
Sigma-Aldrich). The powder was initially subjected to a heat treatment at
800~K for 2~h to minimize the presence of undesired organics and then
subjected to ball milling for 24~h to get a reduced and homogeneous
distribution of particle sizes. Finally, polyvinyl butyral (PVB) was added
as a binder to the powder (3~wt\%) and uniaxially pressed at 150~MPa into
thick bars, followed by isostatic pressing at 250~MPa, and then conventional
sintering at 1350~${^\circ}$C for 2~h. The density measured with Archimede's
method was 5.63~g/cm$^{3}$, 93.6\% of the theoretical density.

For bar \#2, cut with dimensions $43\times 4.1\times 0.59$~mm$^{3}$,
stoichiometric BaTiO$_3$ was prepared by the conventional mixed-oxide powder
technique in the Department of Chemistry of the Martin Luther University
Halle(Saale), Germany: 24~h mixing (agate balls, water) of BaCO$_3$ (Solvay, VL600, $%
<0.1 $~mol\% Sr) and TiO$_2$ (Merck, no. 808), 2~h calcining at 1100~${^\circ%
}$C and after it 24~h fine-milling (agate balls, water). Then the powder was
densified (binder polyvinyl alcohol) to a plate of dimensions $53.5\times
37.0\times 2.0$~mm$^3$ with a density of about 3~g/cm$^3$ and afterwards
sintered 1~h in air at 1400~${^\circ}$C (heating and cooling rate 10~K/min).
The plate was cut into thin bars which were ground down to a thickness of
about 0.6~mm. The relative density was 94\%.

The reducing treatments were carried out in a flux of 0.1 CO + 0.9 Ar at $%
\sim 1$~bar. The samples were put into a long envelope of Pt open at both
ends, with alumina spacers to avoid the direct contact with Pt. The envelope
was inserted into a water-cooled quartz tube where the gas flowed and heated
by induction from outside with a coil connected to a RF power source. Each
reduction treatment lasted $1.5-2.5$~h at temperatures ranging from 950 to
1130~${{}^{\circ }}$C, and was followed by 1~h homogenization at 800~${%
{}^{\circ }}$C in the same reducing flux. The change of $\delta $ was
determined from the mass change with a microbalance. Assuming that the mass
change was entirely due to O loss, the error on $\delta $ was $\sim 10^{-5}$%
. However, the anelastic and dielectric measurements required the
application of Ag paint electrodes, which had to be removed before each high
temperature treatment. During the first part of the research, Ag was
mechanically removed with a blade and emery paper and washed away with
isopropyl alcohol. In this manner, the removal might sometimes be
incomplete, resulting in an overestimation of the O loss from the mass
change, because the evaporated Ag traces would have been counted as O loss.
The error in the determination of each change of $\delta $ is estimated as $%
<10^{-4}$ from the comparison of the cumulative mass losses of sample \#2
and its mass gain after a final reoxygenation, assuming (unrealistically)
that every time the same amount of Ag evaporated. In the end, after
mechanically removing Ag, the samples were further cleaned in a stirred
solution of HNO$_{3}$ in water, $1:9$ in volume ratio, at $50-60$~${%
{}^{\circ }}$C for $2-10$~min, in order to dissolve the Ag traces.\cite%
{OCY98} Reoxygenation was achieved in air at 1100~${{}^{\circ }}$C for 1.5~h
in a Linn High Therm furnace, after which the black samples recovered the
white yellowish color.

The complex Young's modulus $E=E^{\prime }+iE^{\prime \prime }$ was measured
by suspending the bar on two thin thermocouple wires and electrostatically
exciting the flexural resonance. Silver paint was applied to the sample in
correspondence with the exciting/measuring electrode and in order to short
the thermocouple. The measuring system is described in Ref. %
\onlinecite{CDC09}. The real part was measured from the resonance frequency $%
f\left( T\right) $, being $E^{\prime }\propto f^{2}$.\cite{NB72} The data
are presented as compliance $s=1/E$, with $s^{\prime }$ normalized to its
minimum value $s_{0}=s\left( T_{0}\simeq 750~\text{K}\right) $ in the PE
phase; in terms of the resonance frequency, it is $s^{\prime }\left(
T\right) /s_{0}=$ $\left[ f_{0}/f\left( T\right) \right] ^{2}$, where $%
f_{0}=f\left( T_{0}\right) $ is the frequency corresponding to $s_{0}$. The
elastic energy loss, $Q^{-1}=$ $s^{\prime \prime }/s^{\prime }$, was
measured from the decay of the free oscillations or from the width of the
resonance curve.\cite{NB72}

The dielectric permittivity $\epsilon =$ $\epsilon ^{\prime }-i\epsilon
^{\prime \prime }$ was measured by means of a HP~4284A LCR meter with a
four-wire probe and an electric field of 0.5 V/mm, between 10~kHz and 1~MHz,
in a Delta Design 9023 chamber for the temperature control.

\section{Results}

\begin{figure}[tbh]
\includegraphics[width=8.5 cm]{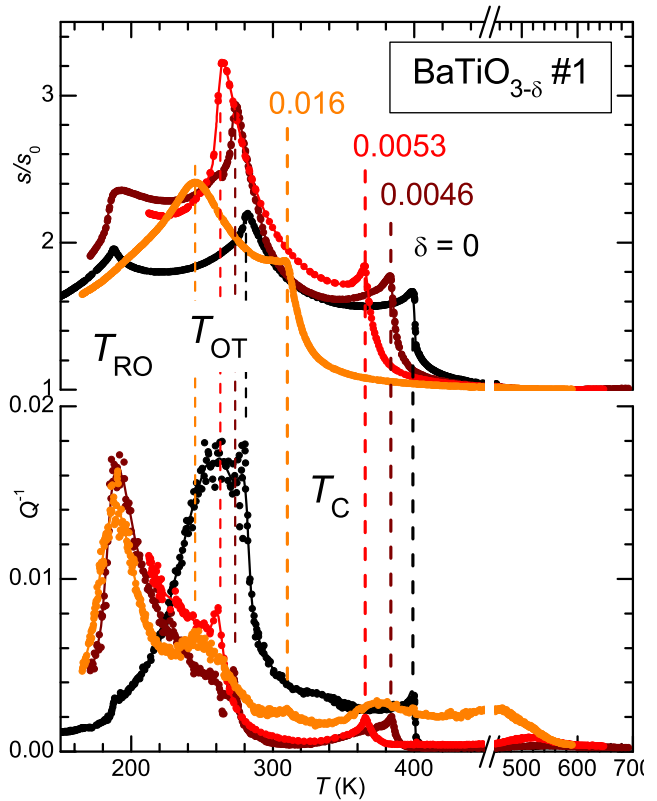}
\caption{Compliance $s$ (reciprocal Young's modulus) normalized to its
minimum value $s_{0}$ in the PE phase and elastic energy loss coefficient $%
Q^{-1}$ of BaTiO$_{3-\protect\delta }$ of type \#1 versus temperature at
various reduction levels. The resonance frequency corresponding to $s_{0}$
ranged between 2.2 and 2.6~kHz. The curves refer to the virgin sample \#1.1 (%
$\protect\delta =0$) and \#1.2 before ($\protect\delta =0.0046$) and after ($%
\protect\delta =0.0053,$ 0.016) reducing its thickness. The Curie
temperatures $T_{\text{C}}$ and the temperatures $T_{\text{OT}}$ of the
transition between tetragonal and orthorhombic states are indicated with
vertical lines. $T_{\text{RO}}$ indicates the transition between the
orthorhombic and the rhombohedral states.}
\label{figL1B}
\end{figure}

Figure \ref{figL1B} presents the normalized compliance $s^{\prime }/s_{0}$
and elastic energy loss coefficient $Q^{-1}$ of BaTiO$_{3-\delta }$ of type
\#1 versus temperature at various reduction levels. The curves are a
selection from various measurements on the two bars \#1.1 and \#1.2: virgin
sample \#1.1 ($\delta =0$) and sample \#1.2 before ($\delta =0.0046$) and
after ($\delta =0.0053,$ 0.016) reducing its thickness. In all cases $f_{0}$
ranged between 2.2 and 2.6~kHz. The Curie temperatures $T_{\text{C}}$ and
the temperatures $T_{\mathrm{OT}}$ of the transition between tetragonal and
orthorhombic states are indicated with vertical lines; $T_{\mathrm{RO}}$
indicates the transition between the orthorhombic and the rhombohedral
phases.\cite{Cor18} In all cases $f_{0}$ ranged between 2.2 and 2.6~kHz. All
the compliance curves presented here were measured during cooling, in order
to avoid any possible influence from aging.

The main effect of doping V$_{\text{O}}$ is a progressive shift of $T_{%
\mathrm{C}}$ to lower temperature, accompanied by an increase of the
steplike softening below $T_{\mathrm{C}}$ ($20\%$ at $\delta =0.016$). Also $%
T_{\mathrm{OT}}$ steadily decreases, and the increase in the peaked
softening at the same temperature is strong ($50\%$ at $\delta =0.0053$),
though at the highest doping it becomes broadened and depressed. The last
transition at $T_{\mathrm{RO}}$ slightly shifts to higher temperature and
noticeably broadens, so that at the highest doping it is visible only as a
peak in the losses.

The losses are peaked at the structural transitions, and exhibit rather high
levels in the FE phases, due to domain wall relaxation. Their general
lowering with increasing $\delta $ can be attributed to the pinning effect
of the V$_{\text{O}}$ on the domain walls, which become static with respect
to the tested frequencies. On the contrary, in the PE phase the losses
increase with $\delta $. This is due to the thermally activated hopping of
isolated and clustered V$_{\text{O}}$, as already found in SrTiO$_{3}$,\cite%
{Cor07} and will be discussed in a separate paper.

\begin{figure}[tbh]
\includegraphics[width=8.5 cm]{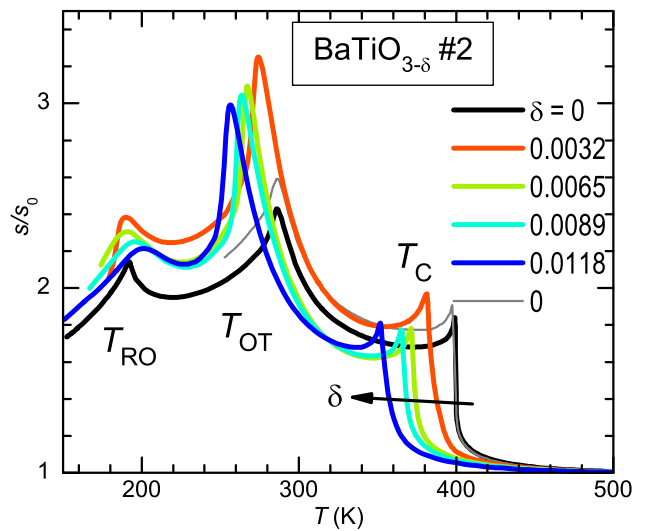}
\caption{Compliance $s$ (reciprocal Young's modulus) normalized to its
minimum value $s_{0}$ in the PE phase of BaTiO$_{3-\protect\delta }$ \#2
versus temperature after various reducing treatments. The thinner curve was
obtained after a final reoxygenation. The resonance frequency corresponding
to $s_{0}$ was 1.85~kHz.}
\label{figBPS0}
\end{figure}

Figure \ref{figBPS0} presents similar results obtained on BaTiO$_{3-\delta }$
\#2. The $Q^{-1}$ curves, which are of no interest here, are omitted for
clarity. The sample was progressively reduced up to a maximum $\delta
=0.0118 $, as estimated from the cumulative mass losses, but after
reoxygenation the mass gain corresponded to $\delta =0.00763$. The actual
discrepancy between the two values, attributable to the evaporation of
residual Ag traces, may be smaller than it appears, since also the last
estimate might have been affected by the evaporation of traces of Ag paint,
this time underestimating the mass gain under reoxygenation. For our
purposes, the exact values of $\delta $ are not important, and the results
in Fig. \ref{figBPS0} are perfectly compatible with those of Fig. \ref%
{figL1B}, in part obtained with a more effective protocol for removing the
Ag electrodes.

After reoxygenation, the anelastic spectrum of BaTiO$_{3-\delta }$ \#2 was
measured again, and found very close to that in the virgin state, except for
an overall 6\% increase of the compliance in the FE phase. The magnitude of
the softening in the FE phase depends on the piezoelectric response of the
material,\cite{Cor18} and a possible explanation for the observed increase
is an improvement in the density, microstructure and domain configuration
after the repeated high temperature reduction and oxygenation treatments.
This fact also excludes any partial conversion of cubic BaTiO$_{3-\delta }$
to the hexagonal phase during the reducing treatments.\cite{RK96} Anyway,
even if some hexagonal phase formed, its effect would be of depressing the
elastic anomalies at the transitions, because its Young's modulus linearly
increases with cooling down to 150~K.\cite{CLM} This would only reinforce
the present observations, that V$_{\text{O}}$ doping does not suppress the
FE transitions, but even slightly enhances the coupling between atomic
displacements and strain.

\begin{figure}[tbh]
\includegraphics[width=8.5 cm]{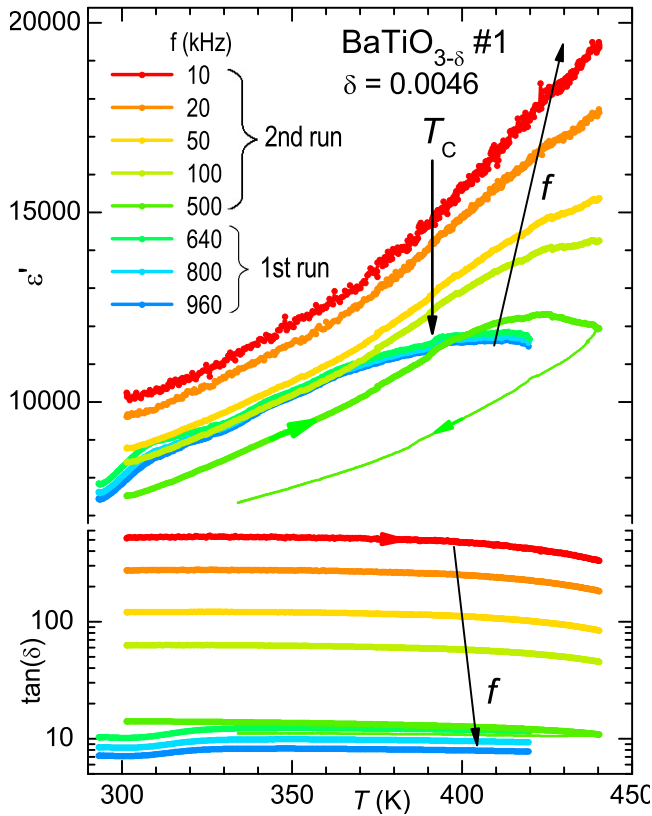}
\caption{Dielectric permittivity and losses of BaTiO$_{3-\protect\delta }$
\#1.2 at the lowest doping, measured during a first heating run at 640, 800
and 960~Hz, and during a second run at 10, 20, 50, 100, 500~kHz. For
clarity, the cooling curves are shown only for 500~Hz.}
\label{fig-diel}
\end{figure}

Figure \ref{fig-diel} presents the dielectric permittivity and losses
measured on BaTiO$_{3-\delta }$ \#1.2 at the lowest doping, $\delta =0.0046$%
. The sample was already black and conductive, though not metallic.

The highest frequencies from 640 to 960~kHz were tested during a first run,
while those from 500~kHz down to 10~kHz during a subsequent run. For
clarity, only the heating data are reproduced, together with the cooling at
500~kHz. These curves have no resemblance at all with the usual permittivity
measured in BT:\cite{Mer49} the sharp peaks and steps at the transition
temperatures are completely absent, $\epsilon \left( T\right) $ steadily
increases with temperature, also above $T_{\mathrm{C}}$, and increases also
with decreasing frequency. The losses are weakly dependent on temperature
and are roughly inversely proportional to $f$ below 100~kHz. These features
are typical of highly conductive materials, and are due to the dynamics of
the free charges. The only effect that can be attributed to the transition
at $T_{\mathrm{C}}$ is a small step in $\epsilon ^{\prime }$ measured at the
highest frequencies, but its sign is opposite to what is expected at a polar
transition.

\section{Discussion}

The first conclusion that one can draw from the present results is that
there is a substantial continuity of the anelastic spectra of BaTiO$%
_{3-\delta }$ with progress of the O deficiency. There is a shift of $T_{%
\mathrm{C}}$ to lower temperature, and some shifting and broadening of the
anomalies at $T_{\mathrm{OT}}$ and $T_{\mathrm{RO}}$, together with changes
in their intensities, but they retain all their characteristics. This
indicates that doping V$_{\text{O}}$ does not induce any major change in the
nature of the three structural transitions, apart from the obvious
introduction of lattice and charge disorder. On the contrary, the dielectric
spectrum in Fig. \ref{fig-diel} is completely changed already at the lowest
values of $\delta $, and does not retain any of the well known
characteristics of the BT dielectric curves, with sharply peaked steps at
the three transitions. This fact, however, is due to the overwhelming effect
of the charge carriers introduced by the ionized V$_{\text{O}}$ defects,
which totally mask any response from the intrinsic electric dipoles. This is
a manifestation of the fact that probing the permanence of polar distortions
in the bulk of a ferroelectric material with high electrical conductivity
cannot be done with the usual methods directly sensitive to the electric
polarization, like dielectric permittivity, hysteresis loop or thermally
stimulated depolarization current, because of the screening effect of the
free charges.

It is then clear that the anelastic spectra provide a practical and valuable
tool for studying highly doped ferroelectrics, in addition to the other 
techniques used so far, notably optical SHG\cite{DLB11,LGP18,WPM17,PWL18} and
Rietveld analysis of diffraction data.\cite{SGW13} The nonlinear optical technique also
allows polar domains to be mapped,\cite{LGP18,DLB11} though it is not
obvious to quantitatively relate the magnitude of the SHG signal with the
spontaneous polarization or piezoelectric coefficients. The determination of
the full structure by diffraction techniques has been exploited\cite{SGW13}
for demonstrating that metallic LiOsO$_{3}$ is isostructural with the LiNbO$%
_{3}$, a well known ferroelectric, and hence a 'ferroelectric-like metal'.
This type of experiment, however, is much more demanding than those
traditionally made in laboratories for FE materials. The measurement of the
elastic moduli as a function of temperature, on the other hand, is
accessible, insensitive to free charge carriers, and the elastic anomalies
at the structural transitions provide insight of the distortion modes
involved in the transitions, and hence indirectly of the atomic
displacements.

Complex compliance curves like those in Figs. \ref{figL1B},\ref{figBPS0}, by
themselves do not tell that the material is ferroelectric below $T_{\mathrm{C%
}}$, since any distortion mode that is coupled quadratically to a strain $%
\varepsilon $ or stress $\sigma $, for example antiferrodistortive
octahedral tilting by an angle $\phi $, would cause a steplike softening
below $T_{\mathrm{C}}$. Yet, when studying the effect of doping on an
initially FE material, it is possible, even without structural measurements,
to figure out if doping, besides introducing charge carriers, may change the
character of the transition from polar to something else. In order to do
that, we discuss the types of transitions other than ferroelectric, that
might possibly be induced by doping in BT or similar FE perovskites. In this
manner, we will discard the possibility that the polar modes of the FE
transitions below $T_{\mathrm{C}}$ are fading away, replaced by other types
of effects induced by doping, and conclude that doping does not depress the
polar modes. Certainly, $T_{\mathrm{C}}$ and also $T_{\mathrm{OT}}$ are
depressed, as one would expect from the lattice and charge disorder
introduced by the V$_{\text{O}}$ and the fact that the induced Ti$^{3+}$
ions, having a radius larger than Ti$^{4+}$, have a reduced tendency to go
off-center. Yet, once the transition occurs, the magnitude of the atomic
displacements and their coupling with strain are not depressed at all, but
rather slightly enhanced. This is especially true for the transition at $T_{%
\mathrm{OT}}$, where the polarization reorients from the $\left\langle
100\right\rangle $ to the $\left\langle 110\right\rangle $ pseudocubic
directions.

\subsection{Elastic anomalies from transitions other than ferroelectric}

Limiting our discussion to perovskites, we should take into account: \textit{%
i)} antiferrodistortive octahedral tilting; \textit{ii)} cooperative
Jahn-Teller; \textit{iii)} antiferromagnetic, ferromagnetic transitions, and
check whether they can possibly appear as a steplike softening below the
transition, so replacing the effect from a fading polar mode.

\subsubsection{Antiferrodistortive octahedral tilting}

The coupling energy between the order parameter for octahedral tilting, the
tilt angle $\phi $, and the involved shear strain $\varepsilon $, is $%
\propto \varepsilon \phi ^{2}$, because rotations of $\pm \phi $ produce the
same $\varepsilon $, and the consequent change in free energy must be the
same. This is the same form of coupling as between spontaneous polarization $%
P$ in a FE and strain, namely the electrostrictive coupling $\propto
\varepsilon P^{2}$, possible also in the PE phase. The simplest situation of
this kind is described by Landau's theory of phase transitions with an
elastic Gibbs free energy\cite{SL98,Cor18c} (with $\sigma $ rather than $%
\varepsilon $ as independent variable)
\begin{equation*}
G=\frac{\alpha }{2}\left( T-T_{\mathrm{C}}\right) \phi ^{2}+\frac{\beta }{4}%
\phi ^{4}-\frac{1}{2}s^{0}\sigma ^{2}-\gamma \sigma \phi ^{2}
\end{equation*}%
where the bilinear coupling $\propto \sigma \phi $ between stress $\sigma $
and order parameter $\phi $ is not allowed by symmetry, and $s^{0}$ is the
compliance in the symmetric high temperature phase. In this case, the
compliance, $s=d\varepsilon /d\sigma $ with $\varepsilon =-\partial
G/\partial \sigma $, remains unaltered until the equilibrium order parameter
$\overline{\phi }$ is null above $T_{\mathrm{C}}$, and is softened of $%
\gamma ^{2}/\beta $ below $T_{\mathrm{C}}$.\cite{ST70,SL98,Cor18c} Sequences
of phase transitions as in BT require additional terms with higher powers of
$\phi $ in the free energy expansion. The point is that, if both the FE
order parameter $P$ and the antiferrodistortive order parameter $\phi $ can
be active, they produce similar types of elastic anomalies and there is no
simple manner to weigh their respective influence on the elastic anomalies.
Such a situation occurs in Na$_{1/2}$Bi$_{1/2}$TiO$_{3}$, where octahedral
tilting and polar modes are mixed in the various phase transitions, and it
is not possible to deduce information on the piezoelectric coupling based
only on the elastic anomalies.\cite{Cor18,Cor18c}

Luckily, in BT there is no tendency at all to octahedral tilting. In fact,
this type of distortion occurs in the ABO$_{3}$ perovskites when the rigid BO%
$_{6}$ octahedra are in compression with respect to the A-O sublattice, for
example when cooling causes a larger thermal contraction in the longer,
softer and more anharmonic A-O bonds, and the octahedra tilt in order to fit
in the smaller volume without distorting the rigid B-O bonds.\cite{Goo04,140}
Barium titanate is the perovskite with the highest ratio of the A-O to B-O
bond lengths,\cite{KN86b} and therefore is the least prone to octahedral
tilting. This is true also after doping with V$_{\text{O}}$, because, even
though Ti$^{3+}$ has an ionic radius of 0.88~\AA , larger than 0.60 of Ti$%
^{4+}$,\cite{Sha76} the V$_{\text{O}}$ weaken the networks of both the
octahedra and Ba-O, and therefore also the strength of the mismatch that
produces tilting. This has been observed\cite{140} in BaCe$_{1-x}$Y$_{x}$O$%
_{3-\delta}$, where a sequence of tilt transitions occurs during cooling:
doping with Y$^{3+}$ on Ce$^{4+}$ increases the average size of the (Ce/Y)O$%
_{6}$ octahedra, because of the larger radius of Y$^{3+}$, and this should
enhance the tilt transition temperature. On the contrary, the temperature of
first tilting transitions is greatly depressed by the larger effect of the V$%
_{\text{O}}$, which weaken the lattice and relieve the mismatch between
octahedra and Ba-O bonds. Similarly, the introduction of V$_{\text{O}}$ in
BaTiO$_{3}$ should not induce any tendency to tilting, in spite of the
increase of the mean size of the (Ti$^{4+}$/Ti$^{3+}$)O$_{6}$ octahedra.

\subsubsection{Cooperative Jahn-Teller transition}

Each V$_{\text{O}}$ in undoped BaTiO$_{3}$ donates up to two electrons,
which are generally localized on Ti, so that a completely ionized V$_{\text{O%
}}$ induces the Ti$^{3+}$ state in two Ti$^{4+}$ ions.\cite{LSH02,LSH04} The
external electronic shell of Ti$^{+3}$ is $3d^{1}$, or an electron is doped
in the empty $3d$ shell. The five degenerate $d$ orbitals are already split
into three $T_{2g}$ ($d_{xy}$, $d_{yz}$ and $d_{xz}$)\ and two $E_{g}$ ($%
d_{3z^{2}-r^{2}}$ and $d_{x^{2}-y^{2}}$)\ orbitals by the octahedral
environment of Ti, which can be further split by octahedral distortions,
giving rise to the Jahn-Teller (JT) effect.\cite{TNY91} The Ti$^{3+}$ ions
are usually observed in acceptor doped titanates,\cite{LSH02,LSH04} where,
at low concentrations, they form polarons or bipolarons, without particular
effects on the structure. It is also suggested that the JT distortions of Ti$%
^{3+}$ in BaTiO$_{3}$ may stabilize the hexagonal phase (h-BT) over the
cubic one under strongly reducing conditions, but this occurs at the very
high sintering temperatures $\sim 1400$~${{}^\circ}$ C, which can be
slightly lowered by doping strongly JT active ions in the Ti site.\cite%
{LMF00b,LMB08} It is out of question that the major lattice rearrangement
necessary for the hexagonal/cubic transition may occur also down to $T_{%
\mathrm{C}}$; if the formation of minor amounts of h-BT occurred in the
present investigation, it was during the reducing treatments at high
temperature. Yet, as explained in the previous Section, this could only
decrease the amplitude of the elastic anomalies, by reducing the fraction of
BT; in fact, h-BT has no phase transitions down to 150~K and above that
temperature its Young's modulus exhibits the usual linear anharmonic
stiffening.\cite{CLM}

Therefore, the only possible effect on the elastic anomalies from the JT
active Ti$^{3+}$ ions would be a tendency to a cooperative JT transition,%
\cite{Goo04} namely ordering of the $3d^{1}$ orbitals and the associated
octahedral distortions into a coherent pattern. Also this scenario could not
explain the observed persistence of the steplike anomaly below $T_{\mathrm{C}%
}$ for at least two reasons. The first reason is that at temperatures as low
as $T_{\mathrm{C}}$ most Ti$^{3+}$ are trapped next to a V$_{\text{O}}$,
which can have up to two axially arranged nearest neighboring Ti$^{3+}$.
These electrons cannot participate to an orbital order transition, because
their occupation, presumably $d_{3z^{2}-r^{2}}$,\cite{LSH04,Kol08} is
determined by the position of the neighboring V$_{\text{O}}$, which can be
considered as static in this respect. Then, the fraction of Ti$^{3+}$ free
from V$_{\text{O}}$ and available for a cooperative JT transition is much
smaller than the maximum possible concentration of Ti$^{3+}$, namely $%
2/3\times \delta $ $\leq 0.01$ in the present investigation, and this is far
below the threshold for cooperative orbital ordering to occur.\cite{Goo04}
Finally, an orbital order transition would cause stiffening rather than
softening,\cite{CTB11,Cor18c} apart from a minor initial softening above the
transition. This is due to the fact that, above the transition temperature,
in the cubic phase, the orbitals are randomly occupied among the $T_{2g}$
and/or $E_{g}$ multiplets, and their populations can almost instantaneously
relax under the influence of the stress from the sample vibration, causing
anelastic relaxation and softening the elastic modulus. Such a softening
disappears below the orbital order transition, where the orbitals are
frozen, appearing as a stiffening.\cite{CTB11}

\subsubsection{Magnetic transition}

The electrons that occupy the Ti$^{3+}$ ions accompanying the V$_{\text{O}}$
carry also unpaired spins, and therefore are potential sources of magnetic
ordering.\cite{CCZ09} Various theoretical papers have been devoted to
predicting possible magnetism in highly deficient BaTiO$_{3-\delta }$, but
the experimental evidence, when not negative, is for very small or
controversial effects.

In BaTiO$_{3-\delta }$ the magnetic susceptibility has been measured at
various reduction levels, characterized with the electron concentration $n$
at 400~K from the Hall-effect,\cite{Kol08} finding a paramagnetic behavior
down to LHe temperature. The paramagnetic rise of $\chi $ below 50~K
persists also at the highest dopings, corresponding to $\delta >0.033$, and
excludes any magnetic ordering at higher temperatures. No magnetism has been
found even in extremely reduced SrTiO$_{3-\delta }$ with $\delta \leq 0.28$,%
\cite{GYN91} while very weak ferromagnetism has been deduced from FC/ZFC
splitting of the magnetic susceptibility,\cite{POL11} but without any
Curie-Weiss peak, under certain reducing conditions and after irradiation
with specific ions and fluences.\cite{POL11} Ferromagnetism has been
reported in nanocrystalline BaTiO$_{3}$, but it disappears after high
temperature annealing with onset of grain growth,\cite{CCW11} so that it can
be ruled out in the present case of well developed grains. In addition,
these effects in the $M-H$ curves are so small to be comparable with those
from usual sources of contamination.\cite{GFV09} In order to obtain more
substantial magnetic responses in BaTiO$_{3}$, one must dope magnetic ions,
but even then the effect of V$_{\text{O}}$ is weak: in EuBaTiO$_{3-\delta }$
the Eu $4f$ spins order antiferromagnetically, and the V$_{\text{O}}$ with $%
\delta $ as high as 0.15 induce a ferromagnetic transition, but only below
3~K.\cite{LZW13}

It can be concluded from the existing experiments that no magnetic ordering
of the spins introduced by V$_{\text{O}}$ is expected at the presently
tested O deficiencies. Even conceding that our samples become strongly
magnetic below $T_{\mathrm{C}}$ under doping, the temperature dependence of
the magnetic contribution to the elastic constants would be different from
the sharp step found at $T_{\mathrm{C}}$. In fact, depending on the strain
derivatives of the exchange constants, there can be either a softening\cite%
{Hau73} or a stiffening\cite{ANS60} below the ferromagnetic transition,
which however is proportional to the square of the magnetization, and
therefore is progressive\cite{Hau73} rather than steplike. Therefore, a
magnetic transition could not replicate the anomaly observed here below $T_{%
\mathrm{C}}$ and even less the peaked softening at $T_{\mathrm{OT}}$.

\subsection{Interpretation of the elastic anomalies in BaTiO$_{3-\protect%
\delta }$ as piezoelectric softening}

After examining the effects of the other types of transitions occurring in
perovskites, we conclude that only octahedral tilting can produce an anomaly
similar to the piezoelectric softening below $T_{\mathrm{C}}$. On the other
hand, due to the large Ba ionic size, stoichiometric BaTiO$_{3}$ is far from
unstable against octahedral titling, and even more after the lattice is
weakened by the introduction of O vacancies. It can be concluded that the
magnitude of the softening below $T_{\mathrm{C}}$ in BaTiO$_{3-\delta }$ is
a measure of the piezoelectric coupling and therefore also of the tetragonal
distortion and the associated off-centering of Ti. In fact, the softening in
the FE phase can be written in terms of the electrostrictive coupling $Q$,
dielectric susceptibility $\chi $ and spontaneous polarization $P_{0}$ as%
\cite{Dev51,CCT16,Cor18c}
\begin{equation}
\Delta s=4\chi _{0}Q^{2}P_{0}^{2}  \label{Ds}
\end{equation}%
ignoring for simplicity the tensorial nature of the various terms. This
softening is also of piezoelectric origin, since in this case the
piezoelectric coefficient is $d=2 \chi_0 Q P_0$.\cite{Dev51,CCT16,Cor18c}
The dielectric susceptibility has been written as $\chi _{0}$, in order to
stress the fact that it is defined as $\chi _{0}=\partial P_{0}/\partial E$,
namely the change of the FE polarization under application of an electric
field. However, in a conducting sample the permittivity $\epsilon \simeq
\chi $ rather measures the response of the free charges to the external
electric field. The free charges screen $E$ outside the sample, and
therefore do not allow the FE dipoles within the sample to feel it. Then,
the appropriate susceptibility to insert in Eq. (\ref{Ds}) is not $\epsilon
^{\prime }$ of Fig. \ref{fig-diel}, but rather the response that the
spontaneous polarization would have in the absence of screening from the
free charges, because the compliance $s$ is measured in the absence of $E$.
None of the quantities in Eq. (\ref{Ds}) can be directly measured in a
conducting sample, but their product is the elastic softening $\Delta s$. If
upon doping $\Delta s$ remains constant or even increases, the most likely
situation seems that all of them vary little. In fact, if one assumes that $%
P_{0}$ vanishes, as commonly implied until recently, at least one or both of
$Q$ and $\chi _{0}$ should correspondingly increase. The electrostrictive
coupling $Q$ is essentially of ionic origin, resulting from the strengths
and geometry of the bonds and ionic effective charges,\cite{SN92} and is
usually weakly dependent on temperature and composition, even through the FE
transition.\cite{LJX14} In the descriptions of FE materials and phase
diagrams, $Q$ is one of the coefficients of the Landau expansion of the free
energy, independent of temperature\cite{Dev51} and sometimes also of
composition\cite{AHB85}. Then, in order to compensate for a hypothetical
vanishing $P_{0}$, one should assume that upon doping $\chi _{0}\sim 1/P_{0}$%
, for which no ground can be found.

We conclude that the enhancement of the softening below $T_{\mathrm{C}}$ and
in correspondence with the subsequent structural transitions reflects the
persistence of ferroelectricity in the presence of free charge carriers,
even in the metallic state. In fact, the critical concentration of electrons
above which BaTiO$_{3-\delta }$ is metallic is\cite{KTK10} $n_{c}=1\times
10^{20}$~cm$^{-3}$, or 0.0064~mol$^{-1}$, which is reached with $\delta
=0.0096$ assuming two electrons per V$_{\text{O}}$. Our highest doping $%
\delta =0.016$ should therefore have induced a metallic state, even assuming
that part of the V$_{\text{O}}$ are aggregated and possibly contribute each
with less than two electrons to doping. These data confirm the initial
finding of the persistence of ferroelectricity in BaTiO$_{3}$ through the
insulator-metal transition, obtained by a combination of resistivity,
specific heat, x-ray diffraction and optical conductivity measurements,\cite%
{KTK10} and demonstrates the effectiveness of studying highly doped or
metallic ferroelectrics by means of elastic measurements.

The results of Kolodiazhnyi \textit{et al.}\cite{KTK10} have been criticized
on the basis of neutron diffraction measurements on BaTiO$_{3-\delta }$ with
$\delta =0.09$ and 0.25, according to which there is a separation into
insulating and metallic phases, rather than coexistence of FE and metallic
states.\cite{JLJ11} Apart from the recent observation of FE in BaTiO$_{3}$
made metallic by La$^{3+}$ doping,\cite{FDO15} there are various replies to
the argument of a phase separation. First, the O deficiencies of the
experiment where the phase separation is found\cite{JLJ11} are far higher
than those we are considering here, and during reductions at those levels
BaTiO$_{3-\delta }$ should, at least partially, transform into the hexagonal
phase.\cite{RK96,LMF00b,LMB08} On the other hand, even if a phase separation
occurred in the present case, the steplike softening below $T_{\mathrm{C}}$
would not shift to lower temperature without reducing amplitude and
sharpness; rather, it would decrease in amplitude with the same $T_{\mathrm{C%
}}$, corresponding to the fraction that remains insulating FE.

Before concluding we comment on the initial increase of the magnitude of the
piezoelectric softening with doping. The increase of the softening below $T_{%
\mathrm{C}}$ is certain at the initial stage of doping, while it might be
influenced by the next transition at $T_{\mathrm{OT}}$ at higher doping, due
its proximity and broadening. Also the transition at $T_{\mathrm{OT}}$ is of
FE nature, consisting in the change of the direction of the spontaneous
polarization from $\left( 001\right) $ to $\left( 011\right) $, and the
associate softening is enhanced by doping even more than that below $T_{%
\mathrm{C}}$; this fact would deserve further investigation but we will
limit to the FE/PE transition below $T_{\mathrm{C}}$. The initial
enhancement of the softening below $T_{\mathrm{C}}$ upon doping might be an
exciting confirmation of the theoretical prediction that the polar
distortions may even be reinforced by metallization, especially if they are
mainly due to local chemical or steric effects.\cite{ZFE18} Yet, some
caution is necessary, since the elastic softening, like the piezoelectric
effect, is affected by the depolarization field $E_{dep}$. The use of Eq. (%
\ref{Ds}) relies on the assumption that the depolarization field is
negligible, a condition that it is argued to be approximated for unpoled
samples probed by vibrations whose wavelength exceed the domain size.\cite%
{CCT16,Cor18,Cor18c} There is however no quantitative assessment in this
sense yet, and it cannot be excluded that at the present conditions $E_{dep}$
partially restrains the piezoelectric softening in the absence of free
charges. In this case, the charge carriers introduced by doping would
neutralize the polarization charges and the associated $E_{dep}$, resulting
in an enhancement of the observed piezoelectric softening. On the other
hand, the magnitude of the piezoelectric softening in these undoped ceramic
samples has been quantitatively compared with the intrinsic piezoelectric $%
\mathbf{d}$ and dielectric constant $\mathbf{\epsilon \simeq \chi }_{0}$
from the literature on single crystals through the appropriate average of
the tensor Eq. (\ref{Ds}), written as $\Delta s=\left\langle \mathbf{d}%
^{+}\cdot \mathbf{\chi }^{-1}\mathbf{\cdot d}\right\rangle $, and good
agreement was found.\cite{Cor18,Cor18c} This fact suggests that in unpoled
ceramics probed at kHz frequencies $E_{dep}$ is indeed negligible in the
majority of domains. In addition, one would expect that the depolarization
charges are already neutralized at the lowest doping $\delta =$ $0.0046$ of
sample \#1, where the mobile charges already totally mask the dielectric
response (Fig. \ref{fig-diel}); instead, Fig. \ref{figL1B} shows that
further doping to $\delta \simeq 0.0053$ still increases the softening below
$T_{\mathrm{C}}$, while we do not consider the highest dopings, where an
influence of the transition at $T_{\mathrm{OT}}$ is possible.

\section{Conclusion}

The topic of coexistence of ferroelectric and highly conducting states is of
both fundamental and practical interest. On the one hand it regards the
fundamental ionic and electronic interactions responsible for
ferroelectricity, and possible new mechanisms of the electron-phonon
interaction, which might, for example, favor superconductivity. On the other
hand there are applications of ferroelectrics, for example in the fields of
thermoelectricity and multiferroics, where the electric conductivity is
either required or an unwanted but frequent byproduct of the material
engineering. In these cases, the major experimental obstacle is probing the
FE state, since the traditional experiments for this purpose apply electric
fields and/or detect electric currents, and hence probe the free charges
rather than the FE response, not to mention the difficulties of poling such
samples.

Here it has been shown that purely elastic measurements are a practical and
effective tool for probing the FE state of materials, independently of their
conducting state. This is possible because in the FE state there is an
additional softening of piezoelectric origin, which can be probed also on
unpoled samples and without the application of electric fields. Therefore,
by following the evolution of the elastic compliance versus temperature at
various doping levels, it is possible to assess how ferroelectricity is
affected by doping. It is also possible to obtain information on unknown FE
transitions in new conducting materials, with an appropriate analysis of the
possible types of structural transitions and couplings to strain.

The method is demonstrated by following the evolution of the Young's modulus
of BaTiO$_{3-\delta }$ versus temperature with $\delta $ up to 0.016, which
corresponds to a metallic state. It turns out that doping shifts the
temperatures of the first two FE transitions, $T_{\mathrm{C}}$ and $T_{%
\mathrm{OT}}$ to lower temperature, but the piezoelectric softening below $%
T_{\mathrm{C}}$ is little affected and even enhanced. After analyzing the
possible alternative mechanisms that might contribute to maintaining the
softened state below $T_{\mathrm{C}}$ under doping through O vacancies, it
is concluded that only the FE transitions originally observed in undoped
BaTiO$_{3}$ can be responsible for the observations. Therefore, doping with V%
$_{\text{O}}$ and the concomitant lattice and charge disorder depress the
onset temperature $T_{\mathrm{C}}$, but not the magnitude of the FE
displacements and their coupling with macroscopic strain. This confirms
earlier conclusions based on a combination of other techniques.\cite{KTK10}

\begin{acknowledgments}
The authors thank P.M. Latino (ISM-CNR) for his technical assistance in the
sample treatments. This work was co-funded by the Coordena\c{c}\~{a}o de
Aperfei\c{c}oamento de Pessoal de N\'{\i}vel Superior - Brasil (CAPES) -
Finance Code 001 (process 88881.062225/2014-01) and S\~{a}o Paulo Research
Foundation FAPESP (grant \#2012/08457-7).
\end{acknowledgments}

\numberwithin{equation}{section}


\bibliographystyle{unsrt}
\bibliography{refs}

\end{document}